# Photon trap for neutralization of negative ion beams

## (The English introduction for the Russian text)

*Popov S.S., Burdakov A.V., Ivanov A.A., Kotelnikov I.A.*

*Institute of nuclear physics SB RAS*

**Introduction**

A traditional approach to produce a neutral beam from the negative ion $H^-,D^-$ beam for further application for plasma heating or neutral beam assisted diagnostics, is its neutralization in a gas or plasma target for detachment of the excess electrons. However, these approaches have a significant limitation on efficiency. At present, for example, for the designed heating injectors with the 1 MeV beam [1] neutralization efficiency in the gas and plasma targets will be 60% and 85% correspondingly [2], which considerably affects the overall efficiency of the injectors. In addition, application of these neutralizers is associated with complications, which might be significant in some applications, including the deterioration of the vacuum conditions due to gas puffing, the positive ions appearance in atomic beam. Photodetachment of an electron from a high-energy negative ions is an attractive method of the beam neutralization. Such method does not require a gas- or plasma-puffing into the neutralizer vessel; it does not produce positive ions and it assists to beam cleaning from fractions of impurity negative ions. The photodetachment of an electron corresponds to the following process: $\boldsymbol{H^- + h\nu \Rightarrow H^0 + e}$. Cross section is well known (see, for example, [3]). It is large enough in a broad photon energy range which practically overlaps all the visible and near IR spectrum. Such photons cannot knock out electron from $H^0$ or all electrons from $H^-$ and produce the positive ions. This approach has been proposed in 1975 by J.H. Fink and A.M. Frank [4]. Since that time a number of projects of photon neutralizer have been proposed. As a rule they are based on an optic resonator similar to Fabri-Perot cells. This needs the very high reflectance mirrors, powerful light source with thin line and very precise tune of all the optic elements. For example, in a scheme considered in [5], the reflectance of mirrors should be not less than 99.96%, the total laser output power is to be 800 kW with output intensity about 300W/cm$^2$ and the laser bandwidth should be less than 100 Hz. It is hardly possible that these parameters together could be realized. In this paper we present an approach based on application of a non-resonance photo-neutralizer. Conceivable characteristics of such a neutralizer are assessed.

# Фотонная ловушка для нейтрализации пучков отрицательных ионов

*Попов С.С., Бурдаков А.В., Иванов А.А., Котельников И.А.*

*Институт ядерной физики им. Г.И. Будкера СО РАН*

*Для эффективной нейтрализации мощных пучков отрицательных ионов водорода и дейтерия давно рассматриваются фотонные мишени. Привлекательность традиционного подхода (эталоны Фабри-Перо) к их созданию ограничивается рядом жестких технических условий и большими экономическими затратами. В данной работе предлагается новая концепция нерезонансной фотонной ловушки (накопителя) для построения технологически более простых оптических нейтрализаторов.*

## I. Введение

Для нейтрализации пучков отрицательных ионов в термоядерных приложениях, таких как нагрев и диагностика плазмы, традиционно используют газовые или плазменные мишени. В таких мишенях происходит «обдирка» (отрыв) избыточных электронов и отрицательные ионы превращаются в нейтральные атомы, которые затем беспрепятственно вводят в замагниченную плазму, тогда как заряженные частицы отражаются магнитным полем. Однако для проектируемых в настоящее время нагревных инжекторов с энергией атомов порядка 1 МэВ для установки ITER [1] такой подход наталкивается на серьёзные ограничения, поскольку эффективность нейтрализации отрицательных ионов в газовой и плазменной мишенях не превышает соответственно 60% и 85%[2][3] из-за того, что существенен сопутствующий процесс полной ионизации до положительного заряда.

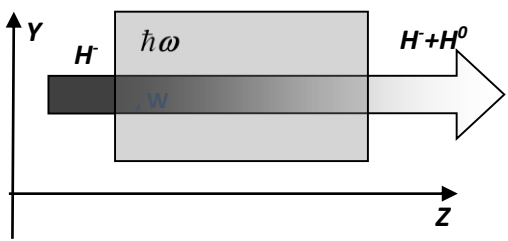

*Рис. 1. Принципиальная схема фотонного нейтрализатора.*

Указанное ограничение существенно снижает КПД проектируемых термоядерных реакторов. Более того, использование газовых и плазменных мишеней сопряжено с дополнительными осложнениями, такими как ухудшение вакуумных условий из-за подвода газа, появление посторонних примесей в пучке и т.п. Например, в проекте ITER предполагается, что примерно 30% отрицательных ионов теряется ещё на стадии ускорения из-за обдирки электронов на нейтральном газе, который проникает в ускоритель из газовой мишени [1] [4].

Перечисленных недостатков лишена фотонная мишень [4] [5] [6] [7] [8] упрощённая схема которой представлена на Рис.1. В такой мишени место газа или плазмы занимает накопитель фотонов. Если энергия фотона превышает энергию сродства электрона атому (0.754 эВ для водорода), но меньше его потенциала ионизации (13.6 эВ), основным процессом взаимодействия пучка отрицательных ионов с мишенью является фотоотрыв электрона от отрицательного иона. Обратный процесс можно подавить, быстро удаляя образовавшиеся электроны, число которых в любом случае заметно меньше числа электронов в плазме или газе. Поэтому в фотонной мишени принципиально можно получить эффективность нейтрализации атомов, близкую к единице.

Нарастание потока (плотности эквивалентного тока) нейтралов в фотонном накопителе описывается уравнением

$$j_0(z) = j_-(0)\left(1 - \exp\left(-\frac{\sigma c}{V\hbar\omega}\int_0^z W dz\right)\right) \quad (1)$$

где $j_-(0) = n_-(0)V$ – плотность потока отрицательных ионов на входе в мишень, $V$ – их скорость, $W$ – плотность энергии газа фотонов, $\sigma$ – сечение фотоотрыва, $z$ – координата вдоль пучка, $\hbar\omega$ – энергия отдельного фотона. Если величину $W$ в формуле (1) считать константой, нетрудно подсчитать, что для мишени с длиной $l$ при плотности энергии

$$W_* = \frac{V\hbar\omega}{\sigma c l} \quad (2)$$

будут ободраны 63% (точнее $1-1/e$) ионов; чтобы ободрать 86% ($1-1/e^2$) или 95% ($1-1/e^3$) ионов, нужно увеличить $W_*$ соответственно в 2 или 3 раза. Сечение $\sigma$ фотоотрыва электрона от отрицательного иона водорода в зависимости от длины волны излучения представлено на Рис.2 по данным работ [9] [10]; теоретически оно вычислено в работе [11]. Поскольку процесс обдирки пороговый, $\sigma = 0$ при длине волны, превышающей 1.64 микрона.

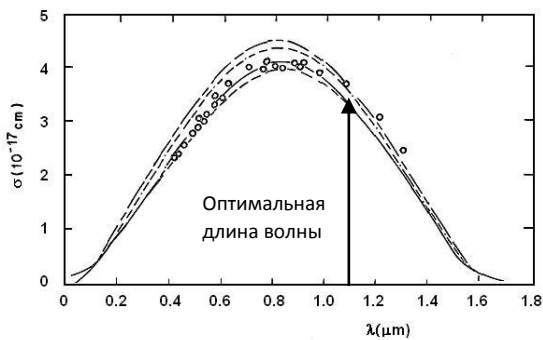

*Рис. 2. Сечение фотоотрыва по данным работ [9],[10],[11].*

Как видно из графика на Рис.2, сечение максимально (а энергетическая цена одного акта нейтрализации минимальна) на длине волны около $\lambda = 1$ мкм. Этот факт в значительной степени определяет выбор источника излучения для фотонейтрализатора. Для примера мы выберем неодимовый лазер Nd:YAG (Neodimium Doped Yttrium Aluminum Garnet) [12] с длиной волны $\lambda = 1{,}064$ мкм ($\hbar\omega = 1{,}24$ эВ, $\sigma \approx 3{,}6 \times 10^{-17}$ см$^2$). При такой длине волны и энергии ионов дейтерия $m_\text{H}V^2/2 = 1$ МэВ для нейтрализатора с длиной $L = 1$ м требуется плотность энергии фотонов $W_* = 5 \times 10^{-6}$ Дж/см$^3$, что соответствует интенсивности излучения $I_* = cW_* = 1{,}53 \times 10^5$ Вт/см$^2$. Достижение столь высоких значений этих величин предполагает большую мощность накачки и/или высокую добротность накопителя фотонов.

Накопитель необходим, поскольку лишь малая доля фотонов поглощается ионами. Как показывает расчёт, выполненный в работе [4] (см. также следующий раздел), в фотонной мишени с параметрами, необходимыми для установки ITER, поглощается лишь 1 фотон примерно из миллиона (кстати, это оправдывает предположение о постоянстве $W$ при вычислении величины $W_*$). По этой причине целесообразно возвращать лазерное излучение с помощью зеркал обратно в область прохождения пучка ионов максимально возможное число раз.

К настоящему времени предложено несколько схем фотонных нейтрализаторов, которые, впрочем, имеют много общего. Все схемы, начиная с первой работы [5], включают мощную систему лазерной накачки и *резонансный* оптический накопитель, который по сути является разновидностью эталона (резонатора) Фабри-Перо. В настоящей работе мы рассмотрим возможность создания *нерезонансного* накопителя фотонов.

Анализу существующих схем оптических накопителей посвящён следующий раздел II. В нем мы рассмотрим некоторые особенности традиционно предлагаемых подходов к данной проблеме и их ограничениям. В III мы предложим альтернативный адиабатический вариант решения, лишенный свойственных резонаторным схемам проблем. В IV в качестве простого примера рассмотрим цилиндрическую геометрию накопителя фотонов. Более технологичная концепция рассмотрена в разделе V с конкретными параметрами для разрабатываемого инжектора в настоящее время [13]. Разделы VI, VII, VIII посвящены сопутствующим техническим проблемам. В IX приведены оценки ожидаемой эффективности работы нерезонансного фотонейтрализатора, далее следует заключение.

## II. Анализ резонансных накопителей фотонов

В работе [6] для накачки оптического накопителя было предложено использовать химический лазер, работающий на сверхзвуковой струе смеси кислорода и паров йода. Длина волны излучения такого лазера составляет $\lambda = 1{,}315$ микрона, а соответствующее сечение фотоотрыва равно $1{,}8 \cdot 10^{-17}$ см$^2$. Тогда для мишени длиной 1 метр и энергии отрицательных ионов водорода 500 кэВ необходимая интенсивность излучения на зеркалах резонатора составит $I_* = 116$ кВт.

Аналогичные предложения, базирующиеся на твердотельных лазерных системах (см., например, [7][8][4] и приведенную в них библиографию) характеризуются аналогичными потоками энергии на стенку резонатора. Несмотря на некоторые отличия оптических схем в этих работах, все они основаны на накоплении фотонов в эталоне (резонаторе) Фабри-Перо, принципиальная схема которого представлена на Рис.3.

В таком резонаторе фотоны накапливаются между двумя параллельными зеркалами с коэффициентами отражения (по амплитуде) $\rho_1$ и $\rho_2$, которые приближаются к единице, а коэффициенты пропускания (по амплитуде) $i\tau_1$ и $i\tau_2$, будучи чисто мнимыми, соответственно близки к нулю.

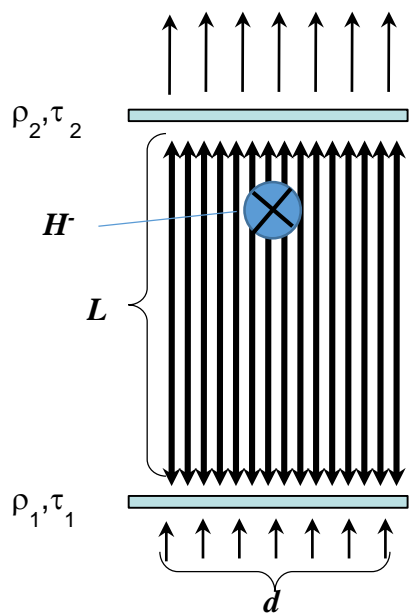

*Рис. 3. Простейшая схема нейтрализатора на основе эталона Фабри-Перо. L – высота области нейтрализации, d – ширина.*

При отсутствии потерь в зеркалах $\rho_1^2 + \tau_1^2 = 1$, $\rho_2^2 + \tau_2^2 = 1$, а поток энергии, падающий снизу на первое зеркало, равен сумме отраженного потока энергии и потока энергии, выходящего из эталона через второе зеркало. Если $\rho_1 \to 1$, почти весь поток, падающий на первое зеркало, от него же и отражается обратно к источнику излучения. Исключение составляют узкие линии в спектре излучения, половина длины волны которых укладывается целое число раз в зазоре между зеркалами. Излучение на такой длине волны, просочившееся в зазор между зеркалами, многократно отражается от зеркал и усиливается вследствие интерференции настолько, что может погасить волну, отражённую от первого зеркала. В итоге кажется, что падающая волна на такой резонансной длине «заходит» в зазор между зеркалами, не отражаясь от первого зеркала. Разумеется, закон сохранения энергии по-прежнему должен выполнятся (до тех пор, пока мы пренебрегаем поглощением в зеркалах и в зазоре), поэтому через второе зеркало из зазора выходит поток, в точности равный падающему на первое зеркало. Поскольку коэффициент прохождения $\tau_2$ через это зеркало мал, ясно, что для равенства входящих и выходящих потоков, амплитуда поля в зазоре должна быть примерно в $1/\tau_2$ раз больше, чем в волне, падающей на это зеркало (то есть сумме всех парциальных волн, циркулирующих между зеркалами), а плотность энергии, соответственно больше примерно в $\zeta \sim 1/\tau_2^2$ раз. Количественно всё сказанное поясняет формула для коэффициента накопления, которую нетрудно вывести, суммируя амплитуды парциальных волн так, как это сделано в учебниках, например, в [14]:

$$\zeta = \frac{c\overline{W}}{I_0} \approx \frac{\mu^2 \tau_1^2 \left(1+\mu^2 \rho_2^2\right)}{\left(1-\mu^2 \rho_1 \rho_2\right)^2 + 4\mu^2 \rho_1 \rho_2 \sin^2 \delta}. \tag{3}$$

Здесь $I_0$ – интенсивность в падающей волны, $\overline{W}$ – средняя плотность энергии в зазоре между зеркалами, $\delta = (\omega/c) L \cos\theta$ – набег фазы волны за один проход между зеркалами при движении луча под углом $\theta$ относительно нормали к зеркалам, $L$ – размер мжзеркального зазора, а $\mu \leq 1$ – амплитудный коэффициент ослабления луча в пространстве при проходе от одного зеркала к другому. Максимальное значение

$$\zeta_{\text{res}} = \frac{\mu^2 \tau_1^2 \left(1+\mu^2 \rho_2^2\right)}{\left(1-\mu^2 \rho_1 \rho_2\right)^2} \tag{4}$$

достигается при резонансных значениях длины волны $\lambda = 2\pi c/\omega$ и угла падения $\theta$, которые определяются из условия $\delta = (2\pi L/\lambda)\cos\theta = \pi n$ при целом $n$. Анализируя выражение (4), предположим, что поглощение в зеркалах отсутствует и в частности $\tau_1^2 = 1 - \rho_1^2$. Тогда легко проверить, что значение $\zeta_{\text{res}}$ максимально, если а $\rho_1 = \mu^2 \rho_2$, т.е. первое зеркало немного прозрачнее чем второе. Соответствующее значение коэффициента накопления

$$\zeta_{\max} = \frac{\mu^2 + \mu^4 \rho_2^2}{1 - \mu^4 \rho_2^2} \tag{5}$$

приближённо совпадает с приведенной выше оценкой $\zeta \sim 1/\tau_2^2$ в пределе $\mu = 1$ и $\tau_2^2 = 1 - \rho_2^2$.

Современные технологии позволяют изготавливать диэлектрические зеркала с коэффициентом отражения $\rho = 0.999$ и выше. Например, в работе [8] заложен коэффициент $\rho = 0.9996$, что соответствует $\zeta_{max} \approx 5 \times 10^3$. Однако, кроме эффективного отражения, в резонансных накопителях необходимо выполнить весьма жесткие требования на качество вводимого лазерного излучения, чтобы выполнялись условия резонанса на большом числе проходов. Если ширина резонанса меньше, чем спектральная ширина лазера накачки, тогда большая часть мощности лазерного излучения отражается от первого зеркала, даже не заходя в резонатор.

Действительно, вне резонанса коэффициент накопления $\zeta$ существенно снижается, причём ширина резонанса тем у́же, чем ближе произведение $\mu^2 \rho_1 \rho_2$ к единице. Полуширина резонанса определяется условиями

$$\frac{(\Delta\theta)^2}{2} = \frac{\Delta\lambda}{\lambda} = \frac{\Delta L}{\lambda} = \frac{\left(1-\mu^2\rho_1\rho_2\right)^2}{4\mu^2\rho_1\rho_2}\frac{\lambda}{2\pi L} = \frac{\left(1-\mu^4\rho_2^2\right)^2}{4\mu^4\rho_2^2}\frac{\lambda}{2\pi L} \ , \qquad (6)$$

которые накладывают чрезвычайно жесткие ограничения на качество источника излучения. Если угловая полуширина $\Delta\theta$, полуширина спектра излучения $\Delta\lambda$ или дрожание зеркала $\Delta L$ превышают указанные значения, накопление фактически невозможно. Для количественных оценок пренебрежём сначала поглощением излучения в зазоре, взяв $\mu = 1$. Тогда при $\lambda = 1$ мкм, $L = 1$ м и $\rho^2 = 0.9996$ получаем $\Delta\lambda/\lambda \leq 2.5 \times 10^{-14}$, что соответствует ширине линии излучения порядка 10 Гц. Создание мощного источника излучения с такой узкой линией является очень сложной задачей. Кроме этого, из-за жестких условий резонанса не менее технически сложна проблема пространственной и температурной стабилизации зеркал.

Коэффициент поглощения излучения в зазоре можно оценить по формуле

$$\mu(z) = \exp\left[-\frac{\sigma j_-(z)}{V} L\right], \tag{7}$$

где $j_-(z) = n_-(z)V$ – плотность потока отрицательных ионов на расстоянии $z$ от входа в мишень. Проект ITER предусматривает создание двух источников нейтральных атомов дейтерия с энергией 1 МэВ или водорода с энергией 870 кэВ с суммарной мощностью 33 МВт. Каждый источник с эквивалентным током в нейтральных атомах $I_0 = 17$А состоит из 4-х сегментов (колонн) сечением $d \times L$ каждая, где $d \approx 10$cm и $L \approx 150$ cm. В [13] сечение пучка практически круглое с характерным размером около 20 см, тем не менее ограничения (6) все равно остаются очень жесткими

В ИТЕР предполагается, что нейтральные ионы получаются из отрицательных ионов в газовом нейтрализаторе с эффективностью 60%. Еще 30% отрицательных ионов теряются в ускорительном тракте из-за обдирки на газе, проникающем в ускорительный тракт из зоны нейтрализатора. Таким образом, источник отрицательных ионов должен иметь ток $I = 17/0.6/0.7 \approx 40$А. Именно эта величина цитируется в проекте ИТЕР [1]. Проектная величина тока отрицательных ионов на входе в нейтрализатор равна $I_- = 17/0.7 \approx 25$А, а плотность тока $j_- = 25/(1.5 \times 4 \times 0.1) \approx 40$А/м$^2$. Для такой величины $j_-$ и $L = 150$см находим $\mu = 1 - 1.4 \times 10^{-7}$ при $\lambda = 1$ мкм. Следовательно, поглощение фотонов в зазоре инжектора с аналогичными параметрами чрезвычайно мало и, вероятно, значительно меньше, чем поглощение в зеркалах, относительно которого у нас нет данных.

Теоретический предел минимально необходимой мощности без учета ограничений (6) мощность лазерного излучения можно оценить по формуле

$$P_* = 4dLI_*/\zeta_{max} = 4d\frac{V\hbar\omega}{\sigma\zeta_{max}}, \tag{8}$$

где $4d = 4 \times 10$ см суммарная ширина одного пучка. Для указанных выше параметров дает сравнительно небольшую величину $P_* \approx 41$кВт, а для [13] $P_* \approx 29$кВт. Однако более полные оценки, учитывающие ряд неизбежных потерь, (см, например [8]) дают существенно более низкое накопление излучения $\zeta = 500$ и, соответственно, значительно большие затраты на излучение.

Подводя итоги анализа схем резонансных накопителей фотонов, можно сделать вывод, что создание фотонной мишени на базе современных технологий в принципе возможно, но требует детальной проработки конструкции накопителя, подбора зеркал и окон для ввода излучения, разработки стационарной излучающей системы необходимой мощности и экономичности. Вопросы стойкости зеркал к бомбардировке различными частицами, связанные с транспортировкой мощного пучка отрицательных ионов в оптическом резонаторе и близости термоядерной плазмы, требуют отдельного рассмотрения. Имеющиеся экспериментальные достижения в создании резонансных накопителей, насколько известно авторам, ограничиваются мелкомасштабными накопителями с размерами порядка $(10 \div 100)\lambda$ и низкой мощностью инжекции [15] [16]. Несмотря на рекордную добротность и число отражений, полученные в этих работах, стоит отметить, что в этих условиях требования по качеству вводимого излучения (6) и пространственной стабилизации сильно облегчаются.

### III. Нерезонансный накопитель фотонов

Предлагаемый нами нерезонансный накопитель фотонов представляет собой систему зеркал, которая обеспечивает многократное отражением лучей и их удержание в заданной области. В отличие от рассмотренных выше резонансных фотонных накопителей, излучение вводится в фотонную ловушку через прозрачное окно в одном из зеркал и далее «запутывается» между зеркалами так, чтобы не выйти обратно через входное окно прежде чем оно поглотится в зеркалах. Трудности реализации фотонного накопителя, таким образом, переносятся в другую плоскость: необходимо воспрепятствовать выходу излучения после недостаточно большого числа отражений от зеркал обратно к источнику через входное окно. Целью последующего изложения в основном является доказательство принципиальной возможности практического решения указанной проблемы.

Представим, что луч света действительно «запутался» в фотонной ловушке. Тогда плотность энергии в этой системе возрастет пропорционально увеличению времени жизни луча в такой ловушке, которое удобно измерять в количестве отражений от зеркал. Как и в резонансных накопителях фотонов это время определяется потерями при отражении от зеркал. Кроме того, луч может выйти за пределы системы как через входное окно, так и через отверстия ввода и вывода пучков частиц. Если пренебречь потерей фотонов через такие отверстия, коэффициент накопления вычисляется по формуле

$$\zeta = \frac{1}{1-\rho^2}. \tag{9}$$

Принципиальное отличие от резонансных накопителей состоит в том, что теперь нет требования фазового синхронизма между большим количеством лучей внутри накопителя. С одной стороны, это уменьшает предельный коэффициент накопления примерно в два раза, но, с другой, существенно облегчается поиск или разработка источника излучения достаточной мощности. Им может быть набор мощных промышленных волоконных лазеров [17],[18].

Рассматриваемая система эквивалентна математическому или световому бильярду. Однако в отличие от хорошо изученных бильярдов Биркхгофа с замкнутыми границами [19], фотонная ловушка является открытым бильярдом, так как лучи могут выходить из неё. Для эффективного нерезонансного удержания необходимо подобрать конфигурацию зеркал, в которой луч испытывает достаточное количество отражений до выхода из системы, которое не меньше чем $N_\rho \sim \frac{1}{1-\rho^2}$. Это возможно, если сохраняются какие–либо точные или приближенные инварианты движения, которые ограничивают область, занятую фотонами. В следующих двух разделах мы рассмотрим несколько схем нерезонансного накопителя фотонов

## IV. Цилиндрическая фотонная ловушка

В некоторых приложениях естественной геометрией для ионного пучка [13] является цилиндрическая. Для фотонной ловушки рассмотрим соответственно аксиальную геометрию. Схема такой ловушки представлена на Рис.4. Она состоит из центральной цилиндрической ячейки с зеркальной внутренней поверхностью и конических концевых отражателей. Как будет показано ниже, при некоторых условиях эти отражатели практически полностью блокируют потери излучения через торцы ловушки.

Будем считать, что излучение волоконного лазера вводится в ловушку через узкое отверстие в цилиндрической части ловушки под углами α и β соответственно к оси и радиусу цилиндра, как показано на Рис.4. Многократно отражаясь от зеркальной внутренней поверхности ловушки оно распространяется вдоль оси ловушки и постепенно затухает вследствие неполного отражения. Величины углов α и β определяют область удержания фотонов в ловушке.

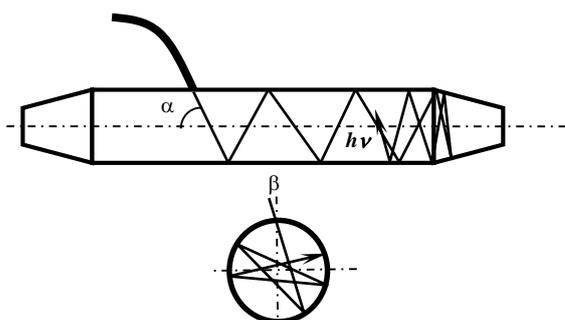

*Рис. 4. Схема цилиндрической оптической ловушки.*

При некоторых значениях α и β траектория луча оказывается замкнутой. Движение фотона по замкнутой траектории является периодическим, и фотон покинет ловушку через входное отверстие после одного периода движения. Такие траектории не подходят для фотонной ловушки, если, конечно, период движения не превышает $N_\rho \sim \frac{1}{1-\rho^2}$. Согласно теореме Пуанкаре о возвращении [20], почти каждая траектория, даже непериодическая, возвращается в свою начальную окрестность. Размер окрестности в приложении к фотонной ловушке определяется диаметром входного отверстия $D$. Необходимо, чтобы время возвращения, выраженное в числе отражений $N$, существенно превышало $N_\rho$.

Нас интересует случай, когда излучение вводится почти перпендикулярно оси ловушки OZ, т.е. $\alpha \approx \pi/2$. Тогда траекторию луча в плоскости, перпендикулярной оси OZ, можно представить «квазимногоугольником» с периметром, пропорциональным диаметру ловушки в плоскости с текущей координатой Z. При $\alpha \approx \pi/2$ достаточно небольшого плавного конического сужения на краях цилиндрической части с малым углом конуса, чтобы эффективно отражать фотоны обратно в цилиндр. При этом угол между осью системы и траекторией луча всё время остается близким к $\pi/2$. Этот факт связан с существованием адиабатического интеграла движения. В самом общем случае эта величина выражается интегралом

$$J = \oint p\,dq \;, \tag{10}$$

вычисленным по циклической координате $q$ в невозмущенной системе, где $p$ – канонический импульс. В рассматриваемом случае невозмущенной системой является круглый цилиндр радиуса $a$, а циклическим – движение по радиусу. Канонически сопряженным импульсом является проекция скорости фотона на радиус вектор $\dot{r} = c\sin(\alpha)\cos(\beta) = c\sin(\alpha)\sqrt{1-a^2\sin^2(\beta)/r^2}$, где $a$ – расстояние от точки отражения до оси.

Вдоль траектории луча, запущенного под углом $\beta$, радиус изменяется в пределах от $a$ до $a\sin(\beta)$. Вычисляя интеграл (10), получаем

$$J = 2 \int_{a\sin(\beta)}^{a} c\sin(\alpha)\sqrt{1 - a^2\sin^2(\beta)/r^2}\,dr = 2ac\sin(\alpha)\left(\cos(\beta) - (\pi/2 - \beta)\sin(\beta)\right). \quad (11)$$

Вследствие аксиальной симметрии на невозмущенной траектории угол $\beta$ сохраняется, поэтому из (11) следует, что адиабатическим инвариантом фактически является произведение $J = a\sin(\alpha)$. Угол $\alpha$ является аналогом питч-угла заряженной частицы при движении в магнитном поле, а сужающиеся конусы на концах цилиндра – аналогом магнитных пробок (см., например,[21]). При уменьшении радиуса цилиндра $a$ угол $\alpha$ должен увеличиваться, чтобы обеспечить сохранение адиабатического инварианта $J = a\sin(\alpha)$. Достигнув значения $\alpha = \pi/2$, луч отражается и начинает двигаться в обратном направлении, удаляясь от концевых конусов. Для сохранения адиабатического инварианта необходимо, чтобы сужение цилиндра на его концах было достаточно плавным.

Как показывает численное моделирование в пакете ZEMAX [22], для удержания фотонов в трубе длинны 200 см и диаметром 40 см с угловым разбросом Δα=±5º относительно перпендикулярной плоскости к оси ловушки достаточно конусов высоты 20-25 см с углом раствора 3º. Траектория луча, испытавшего 1000 отражений, в такой системе

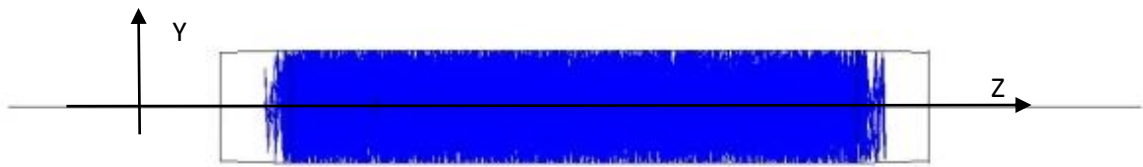

*Рис. 5. Схема ловушки для фотонов с малым продольным разбросом направлений. Синим цветом показан один луч в ловушке со случайным углом инжекции вдоль оси Z (питч-углом) в диапазоне 85-95º и случайным углом в перпендикулярной плоскости(XY) ±20º, испытавший 1000 отражений*

представлен на Рис.5.

Для оценки эффективности удержания вычислялось распределение потока лучистой энергии через продольную плоскость сечения системы по диаметру центральной части. Результат представлен на Рис.6. Пунктирной линией 1 показано положение торца ловушки, 2 – место инжекции излучения. Как видно распределение обрывается, не доходя до торца ловушки. Таким образом, эффективность удержания определяется в основном качеством отражающей поверхности и практически не зависит от качества инжектируемого излучения. Это является главным преимуществом по сравнению с резонансными накопителями (эталоны Фабри-Перо), которые крайне чувствительны к спектральному составу излучения, угловому разбросу, неравномерности тепловой нагрузки зеркал и т.д. [8], [23]

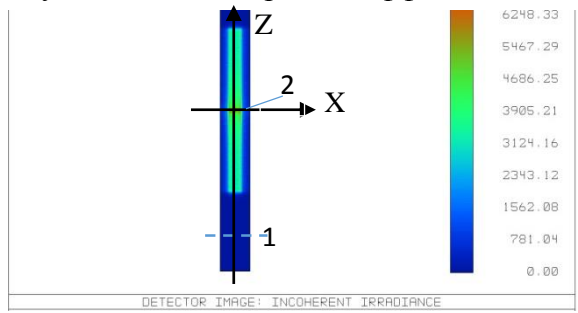

*Рис. 6. Распределение потока лучистой энергии через продольную плоскость симметрии. 1- торец ловушки, 2- область инжекции. Число запущенных лучей – 5000, коэффициент отражения зеркальных поверхностей – 0.995.*

Подбирая диаметр средней части трубы и поперечный угловой разброс инжектируемого пучка, можно сконцентрировать излучение вблизи оси трубы с заданной шириной радиального распределения. Это весьма удобно, поскольку позволяет отодвинуть отражающие стенки от области, занятой пучком отрицательных ионов. На Рис.7 представлены поперечные срезы распределения потока мощности через продольную плоскость симметрии в середине и вблизи конца ловушки. Как видно из рисунка, форма поперечного профиля сохраняется, убывает лишь амплитуда, пикированная в области инжекции.

Проведенное моделирование на простом примере осесимметричной системы показывает принципиальную возможность создавать оптические ловушки с

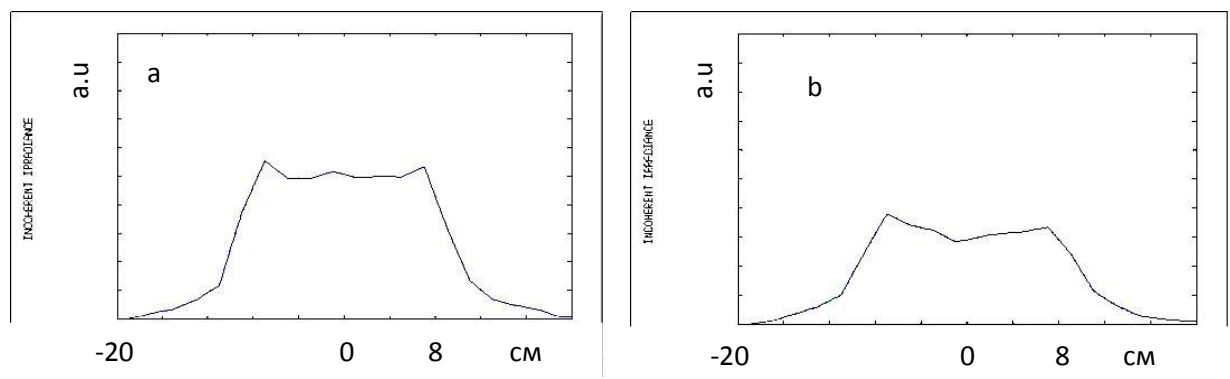

*Рис. 7. Срезы распределения потока мощности в близи центра (a) Z=10см и на периферии (b) Z=90 см ловушки.*

адиабатическим удержанием, не требующие точной настройки элементов и инжекторов с выдающимися спектральными характеристиками. Однако для практического изготовления такая конструкция неудобна, поскольку нет разработанной технологии изготовления зеркальных полых цилиндров. Кажущееся очевидным повышение технологичности путем замены круглого сечения многоугольным с плоскими зеркалами проблемы резко ухудшает удержание фотонов (см. ниже, а также [24])

### V. Квазипланарная открытая оптическая ловушка

Достичь существенного технологического упрощения конструкции можно, переходя к почти плоской геометрии удерживающих зеркал. Двумерный пример такой системы показан на Рис.8.

Как видно из рисунка, при каждом отражении от верхнего зеркала фотон получает приращение горизонтального импульса в ту сторону, где расстояние F до нижнего зеркала больше. При малых отклонениях направления движения фотона от вертикали, он будет стремиться к центральному положению «равновесия». Рассмотрим движение фотона вдоль горизонтали в квазипланарной оптической ловушке для определения условий устойчивости и границ движения фотона. Зададим положение фотона сразу после n-го отражения абсциссой точки отражения $x_n$, ее высотой $F(x_n)$ и углом между вертикалью и скоростью фотона $\beta_n$ (см. Рис.9). Тогда горизонтальное движение описывается следующей системой уравнений:

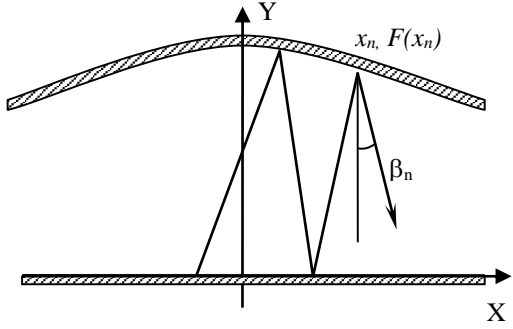

*Рис. 8. Схема квазипланарной фотонной ловушки.*

$$x_{n+1} - x_n = \left(F(x_{n+1}) + F(x_n)\right) tg\beta_n \qquad (12)$$

$$\beta_{n+1} - \beta_n = 2\frac{dF(x_{n+1})}{dx} \qquad (13)$$

Для исследования устойчивости линеаризуем систему (12),(13), получим

$$x_{n+1} - x_n = 2F(0)\beta_n \qquad (14)$$

$$\beta_{n+1} - \beta_n = 2\frac{d^2F(0)}{dx^2}x_{n+1}. \qquad (15)$$

Комбинируя (14) и (15), получим следующее линейное рекуррентное соотношение:

$$x_{n+2} - 2x_{n+1} + x_n = 4F(0)\frac{d^2F(0)}{dx^2}x_{n+1} = -4F(0)\frac{x_{n+1}}{R}, \qquad (16)$$

где $R$ –радиус кривизны верхнего зеркала. Уравнение (16) представляет тип разностной схемы для колебательной системы с единичным шагом по времени и собственной частотой $\omega_0 = 2\sqrt{\dfrac{F(0)}{R}}$. Очевидно решение представимо в виде

$$x_n = A \cdot q^n, \qquad (17)$$

где *q* – комплексная величина. Тогда для *q* имеем:

$$q_{1,2} = 1 - \frac{2F(0)}{R} \pm \sqrt{\left(1 - \frac{2F(0)}{R}\right)^2 - 1}. \qquad (18)$$

Условие устойчивости разностной схемы – $|q| \leq 1$, откуда, учитывая неотрицательность $\frac{F(0)}{R}$, получим условие «геометрического» удержания фотона

$$F(0) < R, \quad \omega_0^2 < 4. \qquad (19)$$

Оценим влияние нелинейности исходной системы на устойчивость при достаточно больших углах наклона лучей, когда существенно нарушается равенство $tg\,\beta_n \approx \beta_n$, но приращение $\beta_{n+1} - \beta_n$ остается существенно меньше амплитуды изменения угла. В этом случае (16) примет вид

$$x_{n+2} - 2x_{n+1} + x_n = -4F(0)\frac{x_{n+1}}{\cos^2(\beta_{n+1})R}. \qquad (20)$$

Косинус в правой части обеспечивает локальное нелинейное увеличение частоты. Как видно, из Рис.8, наибольшая частота – вблизи равновесия, где фотоны имеют максимальный угол к вертикали. Тогда условие устойчивости нужно усилить

$$F(0) < R \cdot (\cos\beta_{\max})^2 \qquad (21)$$

Поправка может оказаться существенной вблизи границы устойчивости.

Как видно, радиус закругления зеркал имеет ключевое значение для удержания фотонов. Это не позволяет использовать одни плоские элементы, совмещенные под конечным углом.

Рекуррентная система (12), (13) позволяет напрямую выделить интеграл движения

$$\sum_n tg\beta_n (\beta_{n+1} - \beta_n) = \sum_n \frac{2(x_{n+1} - x_n)}{F(x_{n+1}) + F(x_n)} \frac{dF(x_{n+1})}{dx} \qquad (22)$$

В случае достаточной гладкости верхнего зеркала и небольших шагов, таких, что

$$\Delta F \ll F, \quad \frac{dF}{dx} \ll 1, \quad \Delta \beta \ll 1 \qquad (23)$$

интегральные суммы (22) приближенно переходят в $\ln \frac{\cos \beta_0}{\cos \beta} = \ln \frac{F(x)}{F(x_0)}$ или в стандартный адиабатический инвариант

$$F(x)\cos(\beta) = const \qquad (24)$$

При выполнении (21), (23) инвариант (24) ограничивает область, занятую фотонами.

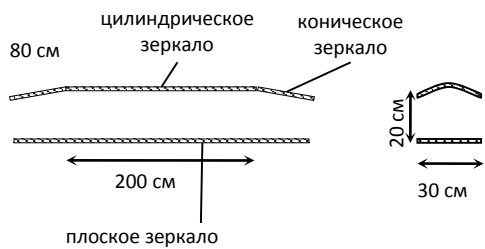

*Рис. 9 Схема квазипланарной адиабатической оптической ловушки для инжектора [13].*

Приведенных оценок достаточно для концептуальной разработки эффективного фотонейтрализатора пучков отрицательных ионов.

Рассмотрим перспективу фотонного нейтрализатора для нейтральной инжекции комплекса Tri Alpha Energy [25]. Из-за особого топливного цикла ($^1$p+$^{11}$B→3$^4$He+8.7MeV) положенного в основу этого термоядерного комплекса приемлемый полный КПД нейтральной инжекции превышает 80%. В связи с этим в разрабатываемый комплекс нейтральной инжекции для Tri Alpha Energy заложена высокая степень нейтрализации пучков H$^-$ (1 МэВ, 10 А). Строящийся прототип такой инжекции представлен в [13].

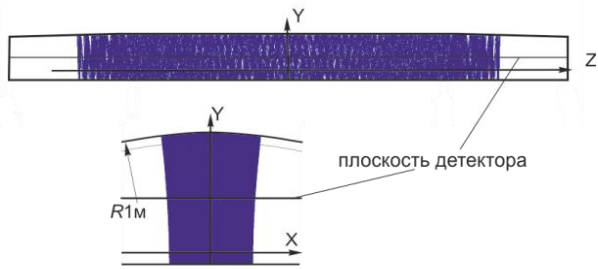

*Рис. 10. Ход одного луча в фотонной ловушке со случайным углом от 0 до 5º в плоскости XY и от 0 до 10º вдоль ловушки, число отражений 4000. Угол раствора концевых конусов около 3º*

Для пучка круглого сечение диаметром около 200 мм в области нейтрализации возможная геометрия ловушки представляет длинную арку из четырех компонентов, как показано на Рис.9: плоского, цилиндрического, и двух конических зеркал. Характерные радиусы закругления элементов с хорошим запасом устойчивости можно выбрать около 1000 мм.

На Рис.10 представлено моделирование хода одного луча в такой системе со случайным углом от 0 до 5º в плоскости XY и от 0 до 10º вдоль ловушки. Представленная траектория содержит 4000 отражений, что дает практически максимально возможную эффективность накопленная мощности излучения при коэффициенте отражения зеркал $\rho^2$=0.999:
$\zeta \approx \dfrac{1}{1-\rho^2} \approx 1000$. Отметим, что в резонансной системе [8] при $\rho^2$=0.9996 авторы полагают получить $\zeta \approx 500$. Потери в [8] прежде всего связаны с большим числом пересекаемых оптических поверхностей на одном проходе и дифракционными потерями.

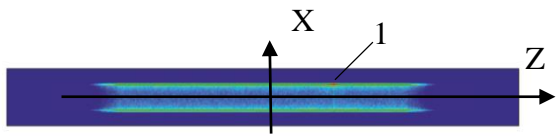

*Рис. 11. Пример поверхностного распределение потока лучистой энергии на детекторную поверхность. 1-точка инжекции излучения.*

Поверхностное распределение потока лучистой энергии в горизонтальной плоскости показано на Рис.11. Детектор обозначен на Рис.10. Коэффициент отражения всех поверхностей взят равным 0.999.

Мощность инжектируемого излучения 1 Вт. Рассчитанная полная мощность на детекторе составила 864 Вт, с учетом цифровых потерь лучей при счете (программа ZEMAX отслеживает и оценивает такие потери) эту величину следует увеличить на 135 Вт. Таким образом, эффективность практически достигает максимально возможной величины. На Рис.12 приведен срез распределения интенсивности излучения на расстоянии 36 см от центра системы. Как видно из рисунков, имеет место пикирование интенсивности вблизи точек поворота излучения, что аналогично распределению концентрации быстрых частиц в открытых плазменных ловушках, например, в установке ГДЛ [26].

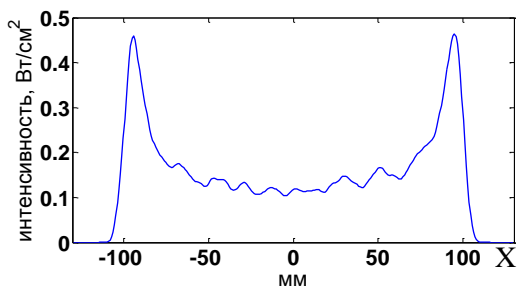

*Рис. 12. Срез поверхностного распределения интенсивности в точке Z=36 см от центра ловушки, X=9см.*

Хотя пики содержат незначительную часть интенсивности, чтобы существенно повлиять на интегральную эффективность нейтрализации, все же профиль можно подправить добавлением источников излучения с другими координатами $X$ и различным угловым разбросом. Результат моделирования комбинации двух источников с половинной мощностью приведен на Рис.13.

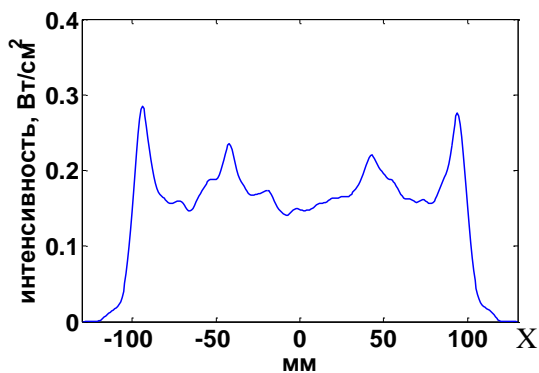

*Рис. 13. Распределение интенсивности в точке Z=36 см, при включении двух источников соответственно с параметрами мощности, положения и углового разброса вдоль X: 0.5 Вт, 0.5 Вт; X=9см, 4 см; 13°, 5°.*

Таким образом, квазипланарная система в пределе геометрической оптики позволяет формировать область удержания с заданными размерами и достаточно резкими границами. Отметим, что концевые конусные зеркала образуют изломы границы с основным цилиндрическим зеркалом. Вообще говоря, это негативно сказывается на продольном удержании фотонов (см выше). Однако, число переходов луча изломов поверхности за время жизни фотона не велико, и он не успевает существенно нарастить продольный угол и вылететь через концы.

В принципе несложно обобщить уравнения (12), (13) на трехмерный случай гладкой поверхности с заданной кривизной верхнего зеркала (это бы теоретически улучшило бы продольную устойчивость), но в данном рассмотрении это не существенно, поскольку продольный масштаб с поперечным сильно отличаются, как и соответствующие периоды продольного и поперечного движения. Из-за этого время вылета фотона продольном направлении, как показывает моделирование, существенно больше времени затухания.

## VI. Дифракционные потери.

Как и в резонаторных схемах запущенный пучок фотонов будет испытывать дифракционную расходимость, накладываемую на геометрический ход лучей. Однако в отличие от первых в рассматриваемых системах дифрагирующие лучи остаются внутри зеркальной системы. Наращивается лишь средний по ансамблю фотонов в пучке разброс углов и, следовательно, область занятая фотонами. В полном объеме рассчитать точные интегралы Кирхгофа на каждом отражении не представляется возможным. Поэтому для верхней оценки увеличения углового разброса можно привести следующие рассуждения. Пусть усредненная по всем отражениям ширина пучка света на верхнем зеркале равна $D$. Тогда дифракционная добавка в угловой разброс на каждом отражении ограничится величиной $\Delta\beta \sim \frac{\lambda}{D}$. С учетом адиабатического ограничения на амплитуду отклонения луча $F(x_{max}) \approx F(0)\cos(\beta_0)$ получим расплывание этой величины и, следовательно, области, занятой лучом $\Delta x_{max} \sim \frac{F(0)\sin(\beta_0)\Delta\beta}{F'(x)} \sim \frac{F(0)\sin(\beta_0)\lambda}{F'(x)D}$. Поскольку $F'(x) \sim \frac{x_{max}}{2R}$, то расплывание составит $\Delta x_{max} \sim \frac{2RF(0)\beta_0\lambda}{x_{max}D}$. При каждом отражении луч расплывается в обе стороны, поэтому, примерно половина фотонов луча уменьшит занимаемую область, а половина увеличит. В этом случае общее расплывание можно считать примерно диффузным, а $\Delta x$ через $n$ отражений оценить как

$$\Delta x_{max} \sim \frac{2RF(0)\beta_0\lambda}{x_{max}D}\sqrt{n} \qquad (25)$$

Характерная ширину пучка света на верхнем зеркале после $n$ отражений с учетом (18) определяется выражением

$$D_n \sim D_0 \sin\left(n \arcsin\left(\sqrt{\left(1-\frac{2F(0)}{R}\right)^2-1}\right)\right), \qquad (26)$$

где $D_0$ -- максимальный поперечный размер пучка внутри одного периода. Очевидно он реализуется в районе центра системы. Отсюда, характерный средний размер пучка света можно оценить, как

$$D \sim \sqrt{\langle D_n \rangle^2} \sim 0.5 D_0 \qquad (27)$$

С учетом (26) и геометрии заданных геометрических размеров ловушки получим $D \sim 0.5\Delta x_0 \sim \Delta\beta F(0) \sim 1\,см$, а для расплывания соответственно

$$\frac{\Delta x_{max}}{x_{max}} \sim \frac{4RF(0)\beta_0\lambda}{x_{max}^2 \Delta x_0}\sqrt{n} \sim 10^{-4}\sqrt{n} \qquad (28)$$

Таким образом, для значимых значений $\beta_0 \sim 0.1$ и $n \sim 1000$, относительное уширение освещенной области будет незначительно: $\frac{\Delta x}{x_0} \ll 1$. Таким образом, дифракционные потери не представляются столь значимой проблемой как в резонаторных схемах.

## VII. Влияние неровностей поверхности

Условие устойчивости удержания фотонов (18) и адиабатичности (23) не допускают на зеркалах дефектных областей с большой кривизной и углом наклона (Рис.9,Рис.11). Потери из-за таких неровностей можно оценить сверху, введя число отражений луча до встречи с такой неровностью

$$j \sim \frac{S}{\Delta S}, \qquad (29)$$

где S – общая поверхность зеркал, $\Delta S$ – площадь дефектных участков. При современных технологиях полировки (см, например, [27]) эти потери являются некритическими.

Технологические проблемы производства таких зеркал требуют отдельного рассмотрения. Очевидно незначительные отклонения кривизны в обработке зеркал большой площади в данной задаче несущественны, как и несущественны отклонения в позиционировании зеркал.

## VIII. Ввод излучения в ловушку. Источники излучения

Для накачки фотонного накопителя авторами предлагается вводить через малое отверстие в зеркале пучок фотонов с заданными угловыми разбросом. Это осуществимо, например, с помощью волоконных лазеров. Действительно, современные технологии предоставляют такие источники. Характеристики одного из промышленных лазеров [28] приведены в *Таб. 1*. Данные серийные лазеры обладают достаточной мощностью, их линия излучения близка к оптимальной. А качество излучения

| Максимальная мощность | 55 кВт |
|---|---|
| $\lambda$ | 1070 нм |
| $\Delta\lambda$ | 6 нм |
| Расходимость | $M^2 \sim 30$ |
| Длина волокна | 30 м |
| КПД | > 30% |
| Размер | 80x280x180 см$^3$ |
| Объем | ~4м$^3$ |

***Таб. 1. Параметры мощного промышленнго волоконного лазера***

достаточно для формирования сходящегося пучка фотонов с требуемым угловым разбросом и ввода его в ловушку через малое отверстие ⌀0.5 мм. Для набора необходимой полной мощности можно использовать несколько лазеров. Ограничение связано только с лучевой стойкостью зеркал в месте первого отражения и потерями на обратном выходе излучения через отверстия ввода. Отношение суммарной площади входных отверстий к площади зеркал должно быть много меньше потерь на одном отражении.

## IX. Эффективность фотонейтрализатора на основе адиабатического фотонного накопителя

Представим степень нейтрализации (1) виде

$$K(P) = 1 - \exp\left(\frac{\sigma P}{E_0 dV}\right), \quad (30)$$

где $d$ - ширина области нейтрализации, $E_0$ - энергия фотона, $V$ - скорость ионов. Интегральная $P$ мощность излучения внутри ловушки равна $P = \frac{P_0}{1-\rho^2}$, где $P_0$-мощность оптической накачки ловушки, $\rho$ – амплитудный коэффициент отражения. Тогда КПД нейтрализации потока отрицательных ионов $P_-$ лазером с КПД $\eta_l$ определяется выражением

$$\eta(P_0) = \frac{K(P_0)P_-}{P_- + P_0/\eta_l} \quad (31)$$

На Рис.14 приведены КПД и степень нейтрализации в зависимости от мощности инжекции фотонов в ловушку. Очевидно, что эффективность растет с ростом мощности пучков H⁻. Для повышения энергетической эффективности можно изменить геометрию пучка H⁻.

Например, осуществлять нейтрализацию несколько круглых пучков, выстроенных в одну линию. Отметим что в этой связи может оказаться очень привлекательным такой подход для нейтральной инжекции ИТЕРа, поскольку там, во-первых, заложен вытянутый профиль пучков, во-вторых, в них планируется использовать дейтерий. В (31) не заложена рекуперация тепловых потерь на генерацию излучения, в случае которой достижимый КПД может быть повышен.

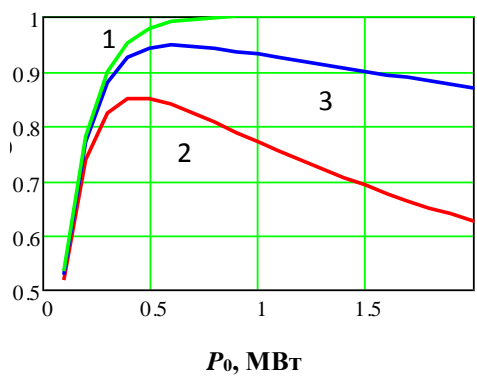

*Рис. 14. Зависимость степени нейтрализации -(1), КПД нейтрализатора при мощности пучка отрицательных ионов 10 МВт, 40 МВт – (2), (3) от мощности инжекции фотонов в ловушку.*

## X. Заключение

В таблице 2 приведено сравнение различных подходов к проблеме нейтрализации. В случае газовой или плазменной мишеней КПД определяется степенью перезарядки, т.к. энергозатраты на работу самих мишеней невелики. В случае рекуперации ненейтрализованных ионов КПД может быть повышен. Тем не менее выходная мощность в нейтралах будет существенно меньше пучка отрицательных ионов в инжекторе. Система с оптической мишенью из-за высокой степени нейтрализации этого недостатка лишена.

В связи с быстрым развитием полупроводниковых лазерных систем не исключено что в будущем станет возможным их использование. Для чего нужно решить проблему угловой расходимости таких источников, или существенно повысить интенсивность. Тогда экономическая эффективность существенно увеличится в следствие более высокого КПД. По-видимому, самую большую проблему в предлагаемой концепции представляет изготовление высокоэффективных зеркал большой площади.

Несмотря на значительную стоимость фотонного нейтрализатора, на большом временном

| Тип мишени | Питание кВт | Оптическая мощность | K | η W=10МВт | η W=40МВт |
|---|---|---|---|---|---|
| Плазменная | ~50 кВт | - | 85% | 85% | 85% |
| Газовая * | - | - | 58% | 58% | 58% |
| Оптический резонатор[4] | 600 кВт | ~200 кВт | 60% | 60% | 60% |
| Оптический резонатор[8] | 3.2 МВт | 800 кВт | 95% | 73% | 88% |
| Ловушка нерезонансная | 1.2 МВт | 400 кВт | 95% | 85% | 92% |

*Таб. 2. Параметры различных нейтрализаторов для пучков отрицательных ионов 10 МВт, 40 МВТ.*

масштабе он может оказаться оправданным по сравнению с плазменными и, особенно, с газовыми нейтрализаторами.

Литература